\pdfoutput=1
\documentclass[]{smule}
\usepackage{url}
\usepackage{hyperref}
\usepackage[colorinlistoftodos]{todonotes}
\usepackage{colortbl}
\usepackage{nicematrix}
\usepackage{amssymb}
\usepackage{amsmath}
\usepackage{booktabs}
\usepackage{adjustbox}
\usepackage{soul}
\usepackage{algorithm}
\usepackage{algorithmic}
\usepackage[inline]{enumitem}
\usepackage{booktabs}
\usepackage{color}
\usepackage{xcolor}
\usepackage{bbding}
\usepackage{listings}
\usepackage{multicol}
\usepackage{xspace}
\usepackage{tablefootnote}
\usepackage{libertinus}

\makeatletter
\AtBeginDocument{

}
\makeatother

\title{Efficient Vocal Source Separation Through Windowed Sink Attention}

\author[1\dagger\ddagger]{Christodoulos Benetatos}
\author[2\ddagger]{Yongyi Zang}
\author[2]{Randal Leistikow}

\affiliation[1]{University of Rochester}
\affiliation[2]{Smule Labs}

\contribution[\dagger]{Work done during internship at Smule}
\contribution[\ddagger]{These authors contributed equally.}

\abstract{
State-of-the-art vocal separation models like Mel-Band-Roformer rely on full temporal self-attention mechanisms, where each temporal frame interacts with every other frames. This incurs heavy computational costs that scales quadratically with input audio length, motivating chunking and windowing approaches. Through analysis of a pre-trained vocal separation model, we discovered that temporal attention patterns are highly localized. Building on this insight, we replaced full attention with windowed sink attention (WSA) with small temporal attention window and attention sinks. We show empirically that fine-tuning from the original checkpoint recovers 92\% of the original SDR performance while reducing FLOPs by 44.5×. We release our code and checkpoints under MIT license at \url{https://github.com/smulelabs/windowed-roformer}.
}

\begin{document}

\maketitle

\section{Introduction}
Vocal source separation is a fundamental task in music information retrieval and audio processing, enabling applications ranging from remixing and music production to accessibility tools and content analysis~\citep{liu2009review}. While state-of-the-art models have achieved impressive separation quality and are widely deployed across the industry, improving their computational efficiency remains an important objective for reducing infrastructure costs and enabling deployment on edge devices.

Current approaches to music source separation fall into three categories: time-domain methods (e.g., Demucs~\citep{defossez2019demucs}, Wave-U-Net~\citep{stoller2018wave}), spectrogram-domain methods (e.g., Spleeter~\citep{hennequin2020spleeter}, Open-Unmix~\citep{stoter2019open}, MelBand-Roformer~\cite{wang2023mel}), and hybrid approaches (e.g., Hybrid-Demucs~\citep{rouard2023hybrid}). Among these, spectrogram-domain methods have emerged as the dominant paradigm, as they operate on lower-dimensional time-frequency representations that facilitate more effective feature extraction compared to the high-dimensional raw waveforms, consistently outperforming alternatives on standard benchmarks.

MelBand-Roformer (MB-R)~\citep{wang2023mel} and BandSplit-RoFormer (BS-R)~\citep{bsr-lu2024music}, both proposed by researchers at ByteDance, represent state-of-the-art architectures for vocal separation. BS-R employs a band-split module to project the input complex spectrogram into sub-band-level representations. MB-R extends this approach by adopting a mel-scale frequency mapping scheme with overlapped sub-bands. After mapping from the STFT domain to mel-bands, both architectures employ transformer layers that process information across time and frequency axes, followed by multi-band mask estimation. This superiority is reflected in the MVSEP leaderboard~\footnote{\url{https://mvsep.com/quality_checker/multisong_leaderboard?sort=vocals}}, where MB-R and BS-R variants, along with their ensembles, dominate the top positions for vocal separation.

However, their reliance on full self-attention mechanisms incurs significant computational costs that grows with input audio length, due to the quadratic scaling nature.\footnote{We note that the quadratic computational complexity is often well-mitigated by the non-autoregressive nature of vocal separation architectures, thereby allowing the operations to be highly parallel and compute faster than in decoder-only architectures.} Motivated by recent studies on sparse attention patterns~\citep{beltagy2020longformer, gao2024seerattention}, we investigated the redundancy in the temporal transformer for MB-R architecture. Specifically, we analyzed the attention patterns of two top-performing checkpoints from the community: the best MB-R and the best BS-R variants. To our surprise, our analysis reveals that temporal attention in both models is predominantly local, suggesting that the full attention mechanism may be unnecessarily expensive. We empirically demonstrate that without any retraining, directly converting the full attention to localized attention yield minimal performance drop when window size is 200 (allowing for 2 seconds of receptive field), and still yields listenable results even when window size drops to 10 (0.1 second receptive field) with an SDR of 5.52 dB, at 80x floating point operations per second (FLOPs) reduction.

Based on this observation, we propose a windowed-attention modification strategy that dramatically reduces computational requirements while preserving separation quality. We implement an efficient FlexAttention-based kernel for windowed attention with sink, and directly retrain the model through a combination of audio-level reconstruction loss and mask-level distillation loss. We empirically validate our approach on the leading open-source MB-R checkpoint, and show that our model can achieve comparable (<1 dB) SDR loss while achieving a 44.5x reduction in attention computations when assessed on 8-second long inputs. We open source our model code and checkpoint, alongside the PyTorch FlexAttention implementation under MIT License at \url{https://github.com/smulelabs/windowed-roformer}.

\section{Related Work}
\label{sec:model-arch}

For the complete architectural details of MB-R, we refer readers to the original papers~\cite{bsr-lu2024music, wang2023mel}. Here we briefly describe the key components relevant to our attention analysis.

\begin{figure*}[h!]
  \centering
\includegraphics[width=\textwidth]{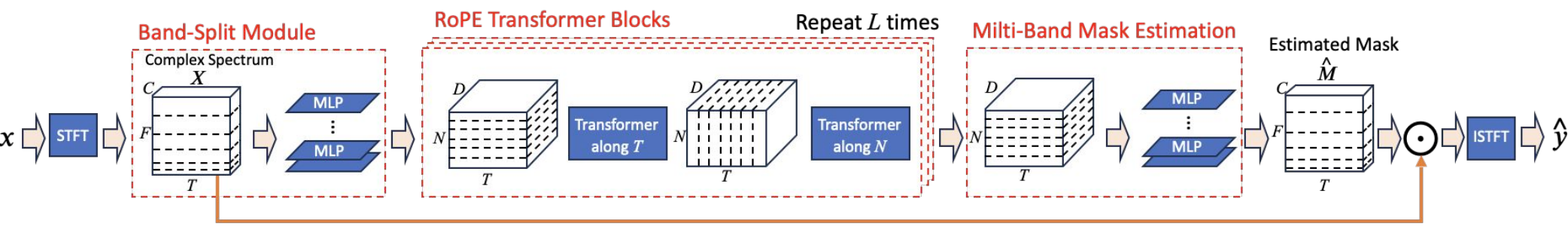}
\caption{The framework of both band-split and mel-band roformer architectures. From \citep{wang2023mel}.}
  \label{fig:model-arch}
\end{figure*}

The MB-R architecture processes audio through the following pipeline: First, a band-split module maps the complex spectrogram to overlapping mel-scale sub-bands. The only difference between BS-R and MB-R is the band scale; MB-R employs a mel scale, whereas BS-R employs a custom splitting scale. Next, RoPE~\citep{su2024roformer} Transformer blocks apply alternating time and frequency transformers with rotary position embeddings. Finally, a multi-band mask estimation module produces complex masks for source separation.

The core of MB-R consists of $L$ transformer blocks ($L=6$ in our analyzed checkpoints), where each block contains two transformer layers. Time transformers process temporal sequences within each frequency band independently, producing attention matrices of size $T \times T$ where $T$ is the number of time frames (801, for 8-second inputs at 10-millisecond hop size). Frequency transformers model inter-band relationships at each time step, generating $N \times N$ attention matrices where $N$ is the number of mel-bands (60 in our configuration). This alternating architecture employs full self-attention in both dimensions, designed to capture arbitrary long-range dependencies in time and frequency.

\section{Methods}
\label{sec:methods}
\subsection{Attention Pattern Analysis}
\label{sec:attn-analysis}

\begin{figure*}[h]
\centering
\includegraphics[width=\textwidth]{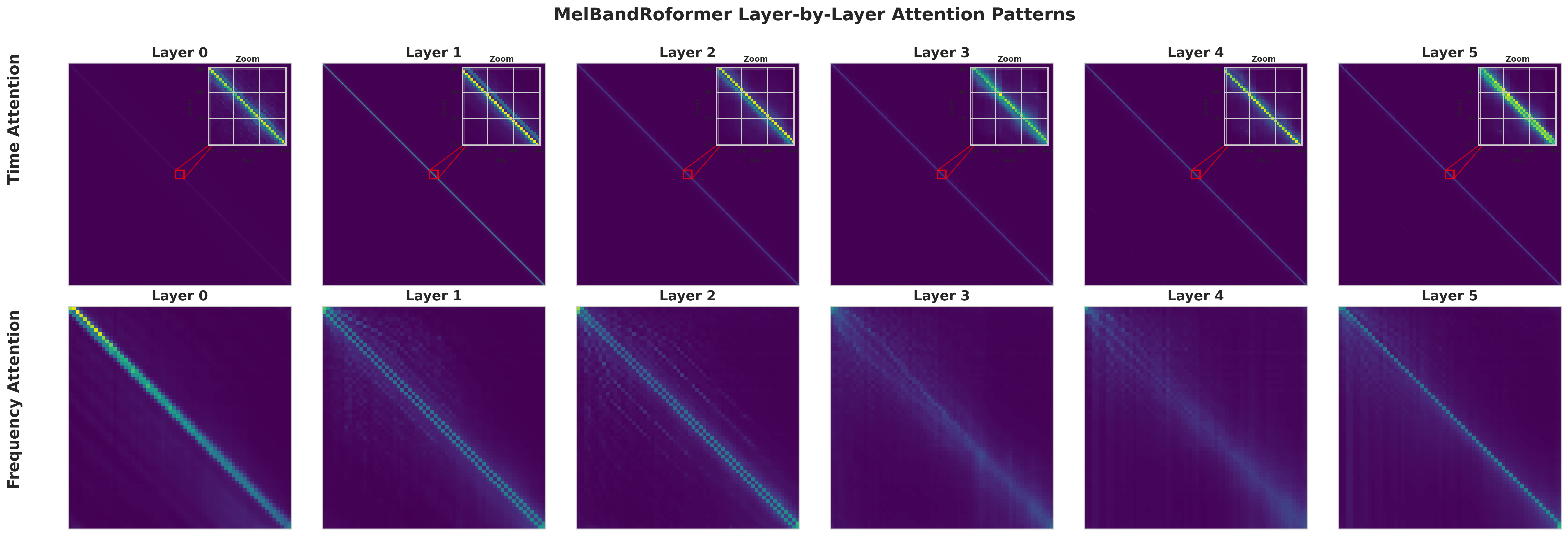}
\caption{Attention patterns in MelBandRoformer transformer layers. \textbf{Top row: } temporal attention shows highly localized patterns concentrated near the diagonal across all layers. We highlight local structure \textbf{(top right)} through 30x30 zoom-in windows. \textbf{Bottom row: }Frequency attention shows more distributed patterns especially in the lower-mid mel bands. BS-R patterns are similar and omitted for space.}
\label{fig:mbr-attention}
\end{figure*}

To identify opportunities for computational optimization, we analyzed attention patterns from two top-performing publicly available MB-R\footnote{https://github.com/KimberleyJensen/Mel-Band-Roformer-Vocal-Model} and BS-R\footnote{bs-roformer-1297 by viperx: https://bascurtiz.x10.mx/models-checkpoint-config-urls.html} checkpoints.
We used the implementation from \citep{solovyev2023benchmarks}. The MB-R checkpoint contains 6 transformer blocks, resulting in 12 total attention matrices (6 time-axis, 6 frequency-axis), while the BS-R has 12 transformer blocks.
We extracted attention maps from 100 random 8 second audio segments with high vocal activity from MUSDB18HQ~\citep{rafii2019musdb18}. For each song and transformer layer, we averaged attention weights across all heads to obtain representative patterns. The consistency of patterns across all examined songs confirmed that our findings represent general model behavior rather than song-specific artifacts.
For visualization, we display full 60×60 matrices for frequency attention to assess global connectivity. For time attention, the full 801×801 matrices show only a thin diagonal line with no visible global patterns—all non-local attention weights are near zero. Therefore, we also show 30×30 windows centered on the diagonal to better visualize the local attention structure that actually exists \ref{fig:mbr-attention}

Our analysis reveals a dichotomy between temporal and frequency attention patterns:

\textbf{Temporal Attention}: Exhibits highly localized patterns across all layers, with attention weights concentrated within approximately 5-10 frames (50-100ms) of the diagonal. This locality remains consistent from early to late layers, suggesting that long-range temporal dependencies are not being utilized despite the quadratic cost of full attention.

\textbf{Frequency Attention}: Maintains more distributed patterns throughout the network, with significant off-diagonal weights indicating global inter-band communication. These patterns are crucial for modeling harmonic relationships and cannot be simplified without degrading separation quality.

This observation motivates our approach: the computational bottleneck of time-axis attention (scaling with $801^2$ attention scores) can be addressed through local attention mechanisms, while the relatively small frequency-axis attention ($60^2$ attention scores) should be preserved. Given that time-axis attention dominates the computational cost by over two orders of magnitude, optimizing temporal attention alone can yield substantial efficiency gains.

\subsection{Windowed Sink Attention}
\label{sec:wsa}

Based on our observations, we designed a sparse attention mechanism to efficiently approximate the original attention patterns, building on previous approaches in efficient attention mechanisms~\citep{beltagy2020longformer, zaheer2020big, xiao2023efficient}. We replace the full $N \times N$ attention matrix (where $N=801$ for time dimension) with a \textbf{Windowed Sink Attention (WSA)} pattern that captures both the strong local dependencies and the diffuse global attention noise. We describe the algorithm in Algorithm~\ref{alg:swa}.

Our WSA consists of two components:
\begin{itemize}
    \item \textbf{Local Window:} A banded diagonal mask with window size $W=10$, where each token $i$ attends only to tokens in the range $[i - W/2, i + W/2]$. This directly models the localized attention patterns we observed, restricting the receptive field to approximately 100ms for a single layer.
    \item \textbf{Sink Tokens:} We append $S=8$ sink tokens to the sequence that have bidirectional full attention with all other tokens. These sinks serve as global information hubs, designed to capture and redistribute the diffuse attention weights present in the original model. This mechanism is analogous to \texttt{[CLS]} tokens in BERT but serves to maintain the global context that would otherwise be lost with purely local attention.
\end{itemize}

\begin{algorithm}
\caption{Sliding Window Attention with Sink Tokens}
\label{alg:swa}
\begin{algorithmic}[1]
\REQUIRE Query $Q$, Key $K$, Value $V \in \mathbb{R}^{B \times H \times N \times D}$, Window size $W$, sink tokens $S$
\ENSURE Output $O \in \mathbb{R}^{B \times H \times N \times D}$

\STATE Mask function $M(i,j) = (i < S) \lor (j < S) \lor (|i-j| \leq W/2)$
\STATE Compute attention scores $A = \text{softmax}\left(\frac{QK^T}{\sqrt{D}} \odot M\right)$

\RETURN $O = AV$
\end{algorithmic}
\end{algorithm}

This design reduces computational complexity from $\mathcal{O}(N^2)$ to $\mathcal{O}(N \times (W + S))$, achieving approximately $80\times$ reduction in FLOPs for our configuration.

To validate this approach, we conducted zero-shot experiments by directly replacing full attention with WSA in pretrained checkpoints:

\begin{table}[h]
    \centering
    \caption{Zero-shot performance and FLOPs reduction for various window sizes on the test set of MUSDB18HQ.}
    \label{tab:zero_shot_wsa}
    \begin{tabular}{lcc}
        \hline
        \textbf{Window Size} & \textbf{SDR (dB)} & \textbf{FLOPs Reduction} \\
        \hline
        Full (801) & 12.12 & $1\times$ \\
        $W=200$ & 11.10 & $4\times$ \\
        $W=100$ & 10.35 & $8\times$ \\
        $W=50$ & 8.71 & $16\times$ \\
        $W=20$ & 6.79 & $40\times$ \\
        $W=10$ & 5.52 & $80\times$ \\
        \hline
    \end{tabular}
\end{table}


While performance degrades with smaller windows, achieving $5.04\text{ dB}$ SDR with $80\times$ fewer FLOPs demonstrates the viability of our approach. The performance gap can be recovered through fine-tuning, where the model learns to redistribute the lost information through the sink tokens.

\subsection{Implementation with \texttt{flex\_attention}}
A naive implementation of WSA would still require materializing the full $N \times N$ attention matrix before masking, negating computational benefits. To achieve practical speedups, we leverage PyTorch's \texttt{flex\_attention} module~\citep{flexattention_blog}, which enables programmatic definition of sparse attention patterns.

The \texttt{flex\_attention} API compiles our attention pattern into optimized CUDA kernels that compute only the specified attention entries, avoiding full matrix materialization. When combined with \texttt{torch.compile()}, this achieves memory and compute efficiency comparable to FlashAttention while maintaining the flexibility to define custom attention patterns. Our implementation reduces peak memory usage from $\mathcal{O}(N^2)$ to $\mathcal{O}(N \times W)$, enabling processing of longer sequences within the same memory budget. We open-source our implementation as part of the model implementation.

\subsection{Training Setup}
\label{sec:training-setup}

We employ \textbf{knowledge distillation} to transfer the learned representations from the original MB-R model (teacher) to our efficient WSA variant (student). Let $\mathbf{x} \in \mathbb{R}^{(B \times C \times T)}$ denote the input mixture and $\mathbf{y} \in \mathbb{R}^{(B \times C \times T)}$ the target source, where $B$ is batch size, $C$ channels, and $T$ time samples. Both teacher $f_T$ and student $f_S$ models produce complex mask estimates $\mathbf{M} \in \mathbb{C}^{(B \times F \times T')}$ after STFT analysis, where $F$ is frequency bins and $T'$ is time frames.

The student model uses WSA with window size $W=10$ and $S=8$ sink tokens. We chose $W=10$ to test the minimal window size that could maintain reasonable performance; larger windows would only improve results but reduce computational savings. The number of sink tokens was chosen arbitrarily and could likely be reduced with further experimentation.

The student model is initialized with the teacher's pretrained weights, ensuring a strong starting point for optimization. The training objective combines two loss components:

\paragraph{\textbf{Reconstruction Loss}}
The primary loss between student predictions and ground-truth targets, using the same $\text{L}_1$ + multi-resolution STFT loss as the original MB-R training:
$$
\mathcal{L}_{\text{recon}} = ||\mathbf{\hat{y}} - \mathbf{y}||_1 + \sum_s ||\text{STFT}_s(\mathbf{\hat{y}}) - \text{STFT}_s(\mathbf{y})||_1
$$
where $\mathbf{\hat{y}} = f_S(\mathbf{x})$ is the student's output and $s$ indexes different STFT resolutions.

\paragraph{\textbf{Mask Distillation Loss}}
We distill knowledge at the mask estimation level $\mathbf{M}$, directly before the inverse STFT operation. 
$$
\mathcal{L}_{\text{distill}} = \lambda_{\text{MSE}} ||\mathbf{M}_S - \mathbf{M}_T||_2^2 + \lambda_{\cos} (1 - \cos(\mathbf{M}_S, \mathbf{M}_T))
$$
where $\mathbf{M}_T = f_T(\mathbf{x})$ and $\mathbf{M}_S = f_S(\mathbf{x})$ are the teacher and student mask estimates, respectively. The MSE term captures magnitude differences while the cosine similarity ensures directional alignment in the complex mask space.

\paragraph{\textbf{Total Objective}}
The total training objective is:
$$
\mathcal{L}_{\text{total}} = \mathcal{L}_{\text{recon}} + \mathcal{L}_{\text{distill}}
$$
We experimented with alternative distillation strategies including layer-wise feature distillation across all transformer blocks. However, mask-level distillation alone achieved the best results, suggesting that constraining only the final learned representations while allowing intermediate features to adapt freely provides the optimal balance.

Training was performed for 200,000 steps on NVIDIA A100 80GB cards on our internal dataset using the original MB-R hyper-parameters from \citep{solovyev2023benchmarks}: Adam optimizer with a constant learning rate of $1\text{e-}5$. The loss weights were set to: $\lambda_{\text{MSE}} = 1.0$ and $\lambda_{\cos} = 1.0$.

\section{Results}
\begin{figure}[h]
\centering
\includegraphics[width=0.8\textwidth]{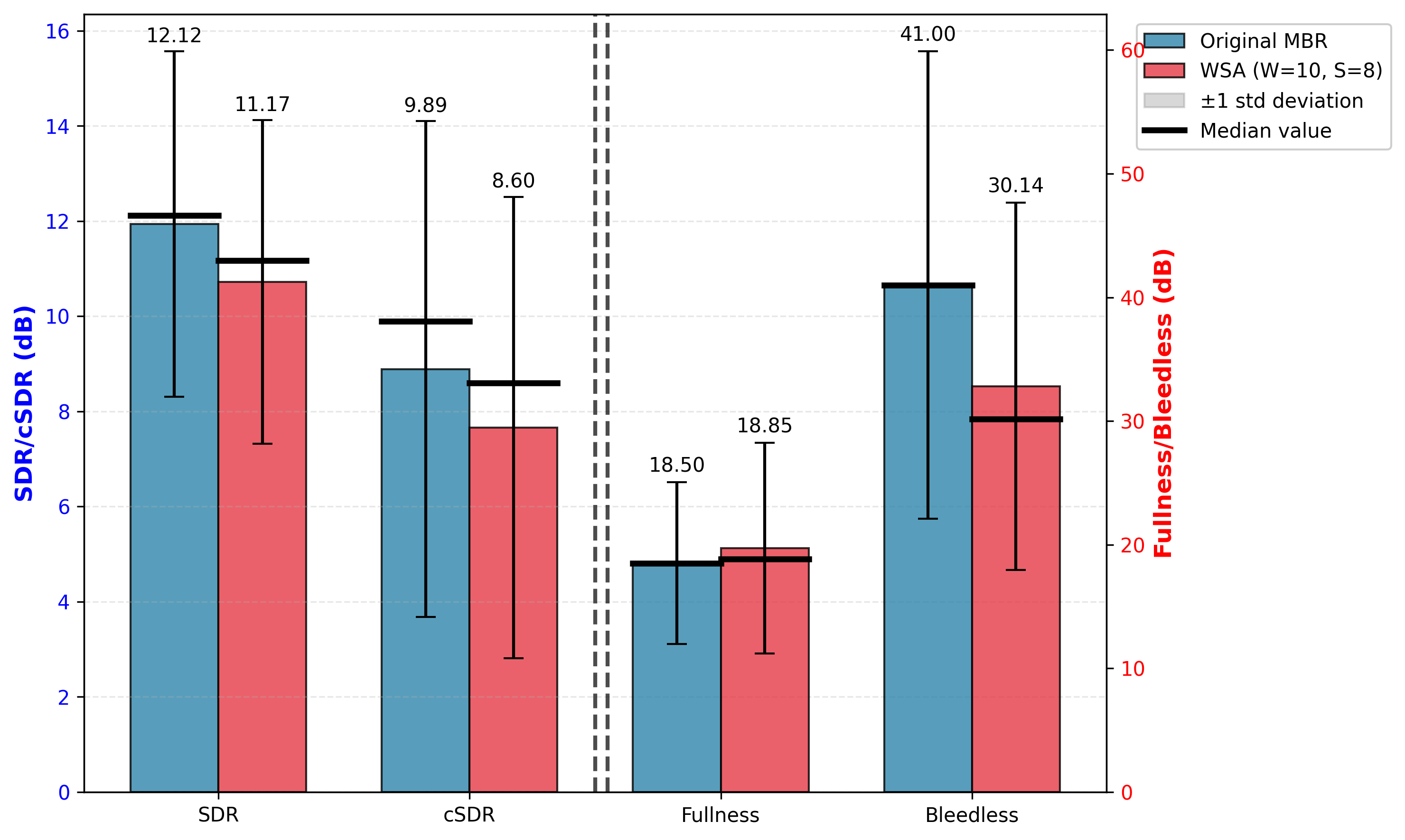}
\caption{Objective testing results of original \textbf{(blue)} and modified \textbf{(red)} variants of MB-R models. Best viewed in color.}
\label{fig:results}
\end{figure}

We evaluated our WSA model against the original MB-R on the MUSDB18HQ test set using multiple metrics. Figure~\ref{fig:results} shows the performance comparison across four evaluation metrics: SDR calculated on complete songs, chunk-based SDR (cSDR) that computes median SDR on 1-second segments, Fullness, and Bleedless.

The WSA model achieves a median of 11.17 dB SDR compared to 12.12 dB for the original MB-R, retaining 92\% (or 0.95 dB less) of the performance. In informal subjective listening, we failed to notice any significant difference in subjective experience between the two model variants, echoing the preserving of original model performance. cSDR shows similar retention with WSA achieving 8.60 dB versus 9.89 dB for the original (87\% retention, or 1.29 dB). We empirically discover that the increase in error primarily occurs in silent segments, where the SDR metric is especially sensitive to small changes in prediction error.

We also evaluated using perceptually-motivated metrics proposed in \citep{solovyev2023benchmarks}. We adopt two metrics: Fullness and Bleedless. \textit{Fullness} measures how completely the target source is preserved by penalizing missing spectral content in the mel-frequency domain, while \textit{Bleedless} quantifies separation quality by penalizing spectral interference from non-target sources. Both metrics are normalized to [0, 100] with higher values indicating better performance. WSA achieves comparable Fullness to the baseline (18.85 vs 18.50), indicating similar preservation of target source content. However, we observe substantially lower Bleedless scores (30.14 vs 41.00), indicating increased interference from non-target sources. Further analysis reveals this degradation is primarily driven by silent segments, where WSA tends to retain more residual content from other sources rather than producing clean silence.

These results validate our hypothesis that temporal attention in MB-R is predominantly local, and our approach in converting to a local processing pattern while retaining performance. By replacing the quadratic full attention (801² = 641,601 operations) with WSA's linear complexity (801×18 = 14,418 operations), we achieve a 44.5× reduction in attention computations while maintaining over 90\% of the SDR performance.\footnote{Results on reported over 8-second segments, as it is the default configuration of the checkpoint.}

\section{Discussions}
\label{sec:discussions}

The practical impact of our 44.5× FLOPs reduction varies significantly across deployment contexts. For the 8-second inference window, modern high-end GPUs can effectively parallelize the original quadratic attention, making wall-clock speedups modest. However, our approach offers compelling advantages in two important scenarios. First, when training on full-length songs (3-5 minutes, or 18,000-30,000 temporal frames), the quadratic scaling of full attention becomes prohibitive. Our windowed attention enables efficient training on complete tracks without chunking artifacts, and we observe substantial wall-clock speedups for sequences exceeding 4,000 frames even on well-provisioned hardware. Second, most consumer devices today have limited parallelism, such as laptops with modest CPUs or entry-level GPUs. On these devices, the computational bottleneck shifts to raw FLOPs rather than memory bandwidth, and our reduction directly translates to faster, more power-efficient inference.

Beyond immediate efficiency gains, our findings reveal something fundamental about temporal structure in music source separation. The strong localization of attention patterns suggests that the inductive biases inherent to audio: tight local dependencies with diffuse global structure along the time axis. This insight is not being fully exploited by current full-attention architectures, and points to a broader opportunity: rather than simply scaling models larger, understanding and leveraging modality-specific structure allows us to push the Pareto frontier of performance versus efficiency. Efficient architectures are not alternatives to scaling but orthogonal enablers of it, making advanced models accessible to researchers without massive infrastructure and allowing those with resources to explore larger design spaces within practical budgets.

\section{Conclusion}
\label{sec:conclusion}

Through analysis of attention patterns in top-performing vocal separation models, we discovered that temporal attention is predominantly local. Based on this insight, we proposed Windowed Sink Attention (WSA), which heavily reduces complexity while preserving essential dependencies through local windows and global sink tokens. Using knowledge distillation with mask-level supervision, our WSA variant recovers 92\% of the original model's SDR performance with 44.5× fewer attention computations on 8-second input segments, and enables better training and inference scaling curves.

\clearpage
\newpage
\bibliographystyle{plainnat}
\bibliography{paper}

\end{document}